\documentclass[letterpaper, 10 pt, conference]{ieeeconf}
\usepackage{color}
\usepackage[english]{babel}
\usepackage[utf8]{inputenc}
\usepackage{amsmath}
\usepackage{graphicx}

\newcommand{\ket}[1]{\bigl| #1 \bigr\rangle}

\title{\LARGE \bf
Wigner function of accelerated and non-accelerated Greenberger–Horne–Zeilinger State}

\author{N. Metwally$^{a,c}$, M. Y. Abd-Rabbou $^{b}$, M. M. A. Ahmed $^{b}$, and  A.-S. F. Obada$^{b}$\\
	$^{a}$ Math. Dept., College of Science, University of Bahrain, Bahrain\\
$^{b}$ Mathematics Department, Faculty of Science, Al-Azhar
		 	University, Nasr City 11884, Cairo\\
		   $^{c}$ Department of Mathematics, Aswan University,Aswan, Sahari 81528, Egypt}

\begin{document}

\maketitle
	\begin{abstract}
The  Wigner function's behavior  of accelerated and non-accelerated  Greenberger–Horne–Zeilinger (GHZ) state is discussed.  For the non-accelerated GHZ state, the minimum/maximum peaks of the Wigner function depends on the distribution's angles, where they are displayed   regularly  at fixed values of the distribution's angles. We show that, for the accelerated GHZ state, the minimum bounds increases as the acceleration increases. The increasing rate depends on the number of accelerated qubits.  Due  to the positivity/ negativity  behavior of the Wigner function, one can use it as an indicators of the presences of the classical/quantum correlations, respectively.  The maximum bounds of the  quantum and the classical correlations depend on  the purity of the initial GHZ state. The classical correlation that depicted by the behavior of Wigner function independent of the acceleration, but depends on the degree of its purity.
\end{abstract}

\section{Introduction.}
Wigner's distribution function  represents one of the closest analogy of quantum mechanics in classical mechanics, where it may be used to  describe quantum mechanical systems by using  a single mathematical objects \cite{PhysRev.40.749}.
Wigner function has many  applications, in statistical mechanics, quantum chemistry,  and  quantum optics \cite{schleich2011quantum}.
 Moreover,  the finite dimension Wigner function is  correlated with quantum information processing via the quantum state tomography, algorithm, teleportation, and estimation of the efficient of quantum circuit\cite{PhysRevA.64.034301,PhysRevA.65.062309,d2003spin,PhysRevA.72.012309,PhysRevLett.115.070501}.
  The positive values of the  Wigner function indicate the classicality of the systems, while the negative values means that the system has quantum correlations \cite{nielsen2000quantum}.   There are many  studies   discussed the behavior of the  Wigner function for different  quantum systems\cite{PhysRevA.48.2479}.
   For qubit systems, there are some limited studies of Wigner function, for example, the time evolution of superconducting flux qubits coupled to a system of electrons is analyzed by the  Wigner distribution\cite{reboiro2015use}.
    For both continuous and spin variables the association between the Wigner and the tomographic probability distribution are discussed in \cite{man2012states}.

Recently, the tripartite   Greenberger– Horne– Zeilinger (GHZ) and W states are reconstructed experimentally by Wigner distribution   \cite{ciampini2017wigner}. In the context of quantum information, the GHZ state has discussed in many detections. As an example,
the quantum Fisher information with respect to SU(2) under some  decoherence channels is obtained analytically \cite{PhysRevA.84.022302}. The entanglement of the  GHZ-symmetric states is discussed in  \cite{PhysRevLett.108.020502}. An efficient scheme to generate the multi-partite GHZ state by using via different techniques is investigated in  \cite{PhysRevA.83.062302,PhysRevA.83.042313}.

   Nowadays, the  Unruh Hawking effect\cite{PhysRevA.82.042332} has discussed for different quantum systems. For example,  M.R. Hwang studied  the entanglement of  a tripartites system \cite{PhysRevA.83.012111}.
    The entanglement of tripartite fermionic system are discussed in non-inertial framework\cite{khan2014tripartite}. N. Metwally, discussed  the non-inertial processing of  a qubit and a qubit-qutrit systems\cite{metwally2016entanglement}.
Therefore, we are motivated to discuss the behavior of the Wigner function for the  accelerated and non accelerated GHZ state, where we assume that the initial system is initially prepared in a non-pure GHZ state. We investigate the effect of the mixing, acceleration parameters  as well as the  distributions angles. 

This paper is organized as following: in Sec.2, we review the derivation of the Wigner function  for the tripartite state.  The behavior of  Wigner function for  the non-accelerated GHZ state is investigated in Sec.2.1.  For the accelerated GHZ state, the Wigner function is discussed in Sec. 2.2. Finally, we summarize our results in Sec.3

 \section{ Wigner function of a tripartite system.}
In SU(2) algebra, the atomic coherent state family of Q-PD   is described by  $s$ parameter function. In the angular  momentum basis it takes the form\cite{PhysRevA.24.2889,PhysRevA.61.034101}:
\begin{equation}\label{4.1.1}
Q^{(s)}_{\hat{\rho}}(\theta,\phi)=Tr[\hat{\rho}_{a}\hat{R}_a^{(s)}(\theta,\phi)],
\end{equation}
where the parameter $ s $ take the values -1, 0 and 1, for the Q-function, the Wigner function, and the P-function, respectively.
The Q-PD for the isolated tripartite system is defined as:
\begin{equation}\label{4.1}
W^{(s)}_{\hat{\rho}}(\theta,\phi)=Tr[\hat{\rho}_{a,b,c}\hat{R}_a^{(s)}(\theta,\phi)\hat{R}_b^{(s)}(\theta,\phi) \hat{R}_c^{(s)}(\theta,\phi)],
\end{equation}
 where the operator $ \hat{R}_i^{(s)}(\theta,\phi) $ is defined as:
\begin{equation}\label{4.2}
\hat{R}_i^{(s)}(\theta,\phi)=\sqrt{\frac{4 \pi}{2J+1}} \sum_{L_i=0}^{2J} \sum_{K=-L}^{L} (C^{J,J}_{J,J;L,0})^{-s} \hat{T} ^{(i)^{\dagger}}_{L,K} Y^i _{L,K}(\theta,\phi),
\end{equation}
 where $ J=N/2$, $ 0 \leq L \leq 2J $ and $Y^i_{L,K}(\theta,\phi) $ are the spherical harmonics, while the coefficient $ C^{J,J}_{J,J;L,0} $ is the usual Clebsch-Gordan for the angular
 momentum of size $ J $. The kernel operator $  \hat{T} ^{(i)^{\dagger}}_{L,K} $ is the irreducible tensor operator (ITO) which is defined in the standard angular momentum basis $ |k,J\rangle  , k=-J,...,J $ as follows:
 \begin{equation}\label{4.3}
 \hat{T}^{(i)^{\dagger}}_{L,M}=(-1)^{M}\sqrt{\frac{2L+1}{2J+1}} \sum_{k,k'=-J_i}^{J_i} C^{J_i, k'}_{J_i,k;L,-M}|J_i,k'\rangle\langle J_i,k|,
 \end{equation}
  where $ -L\leq M \leq L $. By setting  $ J=1/2 $ we can calculate the $s$-parameterized of the Q-PD for  as a follow:
  \begin{eqnarray}\label{4.4}
  W^{(s)}_{\hat{\rho}}(\theta,\phi)&=&(2\pi)^{\frac{3}{2}} \ Tr \bigg[\prod_{i=a,b,c}^{}\hat{\rho}_{i}\big( \hat{T}^{i^{\dagger}}_{0,0} Y^i_{0,0}(\theta,\phi)
  \nonumber\\
  &&+ (\sqrt{3})^{(s)} \sum_{n=-1}^{1} \hat{T}^{i^{\dagger}}_{0,n} Y^i_{0,n}(\theta,\phi)\big)\bigg],
  \end{eqnarray}
  where,
  \begin{eqnarray}
  \hat{T}^{(i)^{\dagger}}_{0,0}&=&\frac{1}{\sqrt{2}}(|0\rangle_i \langle 0|+ |1\rangle_i \langle 1|), \quad
  \nonumber\\
  \hat{T}^{(i)^{\dagger}}_{1,0}&=&\frac{-1}{\sqrt{2}}(|0\rangle_i \langle 0|- |1\rangle_i \langle 1|), \quad
   \nonumber\\
   \hat{T}^{(i)^{\dagger}}_{1,-1}&=&|1\rangle_i \langle 0|\quad
   \hat{T}^{(i)^{\dagger}}_{1,1}=-|0\rangle_i \langle 1|
   \end{eqnarray}
 and
\begin{equation*}
	\begin{split}
	|1\rangle= |\frac{1}{2}, \frac{1}{2}\rangle=|\frac{-1}{2}, \frac{-1}{2}\rangle,
	\quad |0\rangle= |\frac{-1}{2}, \frac{1}{2}\rangle= |\frac{1}{2}, \frac{-1}{2}\rangle.
	\end{split}
\end{equation*}

\subsection{The non-accelerated GHZ state}
It is  assumed that, the system is initially prepared in a non-pure GHZ state. In the computational basis, the  GHZ state  is given by, \cite{PhysRevA.63.054301}:
\begin{equation}\label{4.5}
	\hat{\rho}_{GHZ}=\nu|GHZ\rangle \langle GHZ| +\frac{1-\nu}{8} I_8
\end{equation}
where $\nu$ is the mixing parameter and  $ 0\leq\nu\leq1 $, and $ |GHZ\rangle =\frac{1}{\sqrt{2}} (|000\rangle+|111\rangle)$. This state is known to be Bell nonlocal for $ \nu >0.5 $, while it  is separable if and only if $ \nu \leq 0.2 $.

The Wigner function of the  non accelerated GHZ state  is obtained after a few simple calculations from Eq.(\ref{4.4}) and (\ref{4.5}). For simplicity we assume that the spherical harmonics of the three  qubits are independent and equals, i.e. $ Y^a(\theta,\phi)=Y^b(\theta,\phi)=Y^c(\theta,\phi)=Y(\theta,\phi) $. Thus the Wigner function of the non-pure GHZ state is given by,
\begin{eqnarray}\label{4.6}
	W^{(0)}_{\hat{\rho}_{GHZ}}(\theta,\phi)&=&\frac{1}{8} \Bigl[3 \sqrt{3} \nu \sin ^3(\theta ) \cos (3 \phi )
\nonumber\\
&&+9 \nu \cos ^2(\theta )+1\Bigr]
\end{eqnarray}

\begin{figure}[h!]
	\includegraphics[width=0.49\linewidth, height=3.5cm]{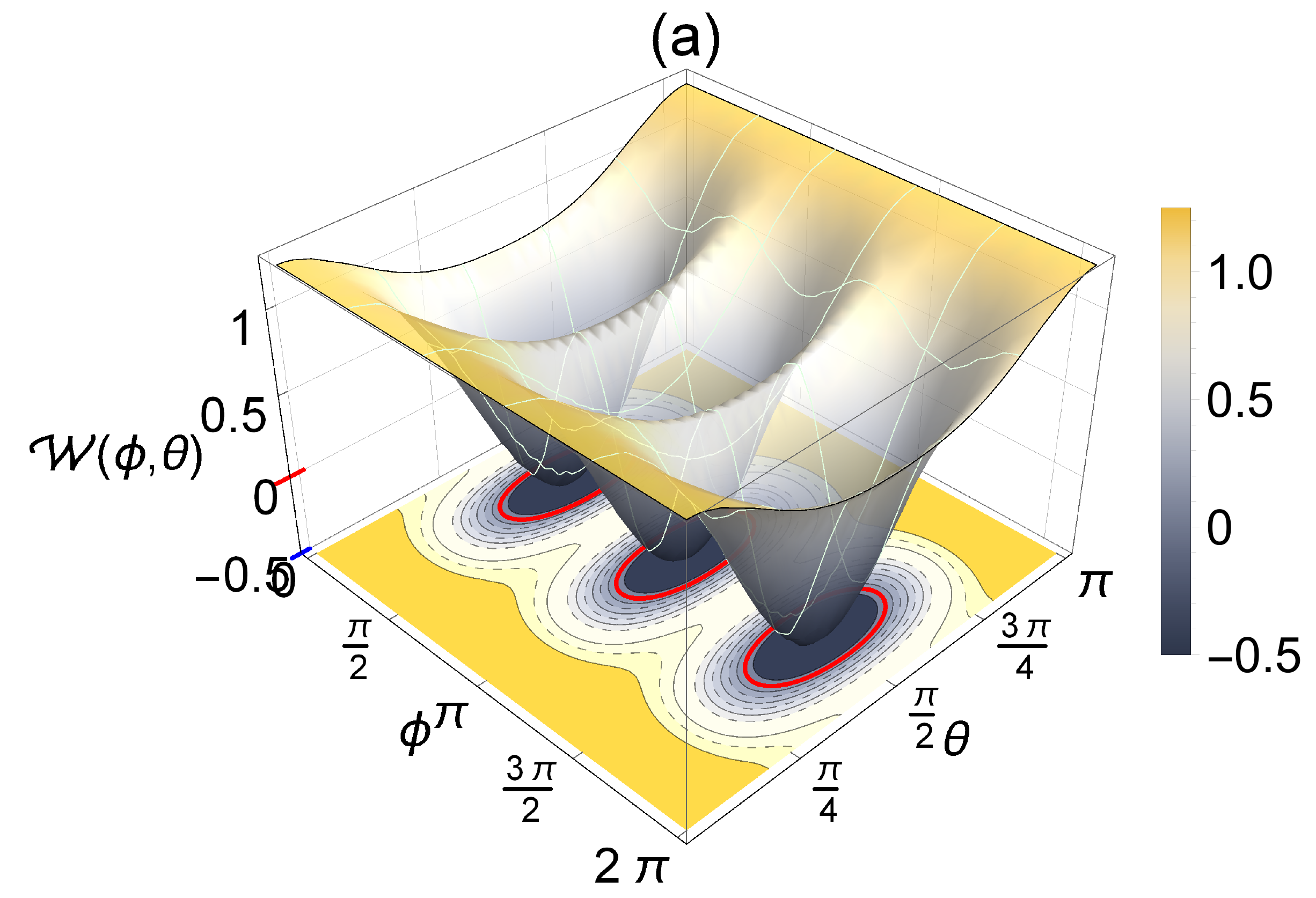}
	\includegraphics[width=0.49\linewidth, height=3.5cm]{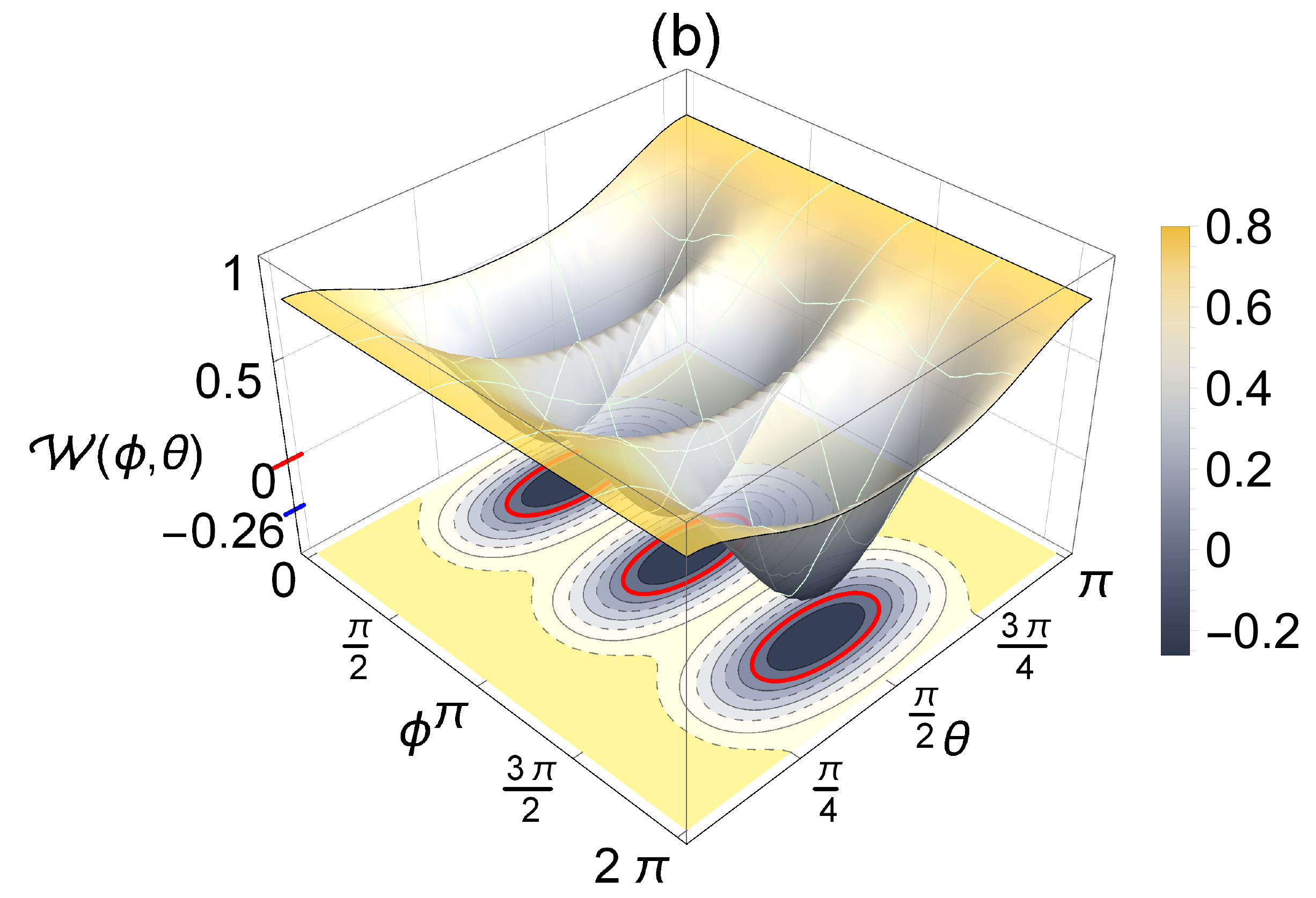}\\		
\includegraphics[width=0.8\linewidth, height=3.5cm]{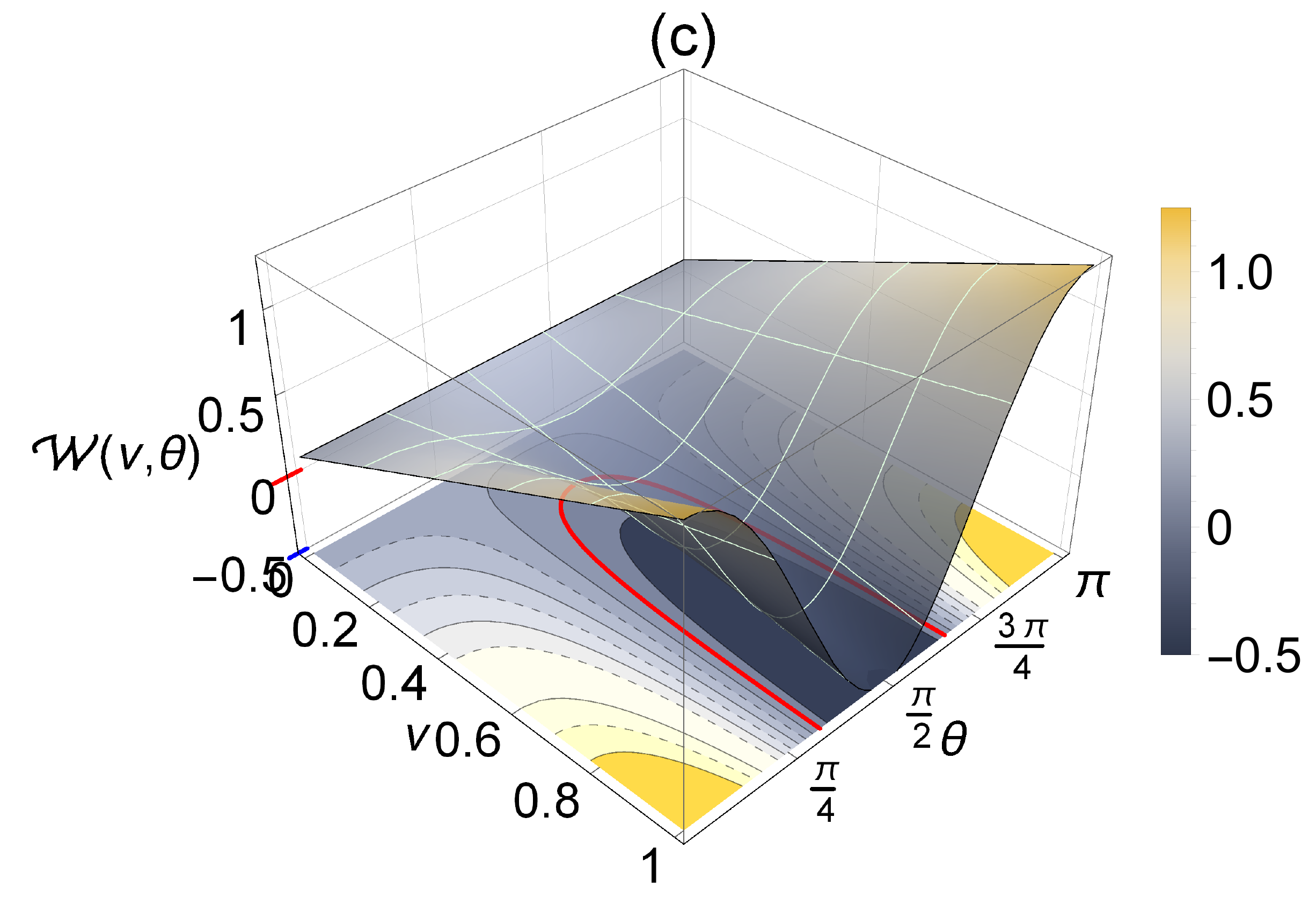}
	\caption{The Wigner function $ W(\phi,\theta) $ of the non-pure GHZ state, at (a)  $\nu=1$,  (b) $\nu=0.3$,  and (c) represents the Wigner function $W(\nu,\theta)$ at $\phi=\pi$ }
	\label{fig:4.1}
\end{figure}

 Fig.(1), displays  the behavior of Winger function  for the non-pure  GHZ state, where different values of the mixing  parameter $\nu$ are considered.
 It is clear taht, theminimum values of$W(\theta~,\phi)$ depend on the purity degree of the GHZ state. As it is shown in Fig.(1a), the minimum bounds of the Wigner function are smaller than those displayed in Fig(b), where we set $\nu=0.3$.
 The  negative regions are centered around $\theta=\pi/2$ and different intervals of $\phi$. The positivity of $W(\theta~,\phi)$, indicates that the it contains classical correlation even its a completely pure state.
    As it is displayed in Figs.(1c), by  increasing  the parameter $\nu$, the minimum bounds $W(\theta,\phi)$  decrease which means that the  quantum correlations increase. Theses minimum bounds are displayed at for any value of $\nu>0.2$ and $\theta\in\simeq[\pi/4,~5\pi/4]$

\subsection{The accelerated GHZ state}

In this section,  three  cases are considered, either one, two or three qubits are accelerated. In the  computational basis $\ket{0_i}$ and $\ket{1_i}$ are transformed from the Minkowski coordinates into Rindler coordinates as a form\cite{PhysRevA.82.042332}:
\begin{eqnarray}\label{4.7}
	|0_i\rangle&=&\cos r |0_i\rangle_{I} |0_i\rangle_{II} + \sin r |1_i\rangle_{I} |1_i\rangle_{II} ,
\nonumber\\
 |1_i\rangle&=& |1_i\rangle_{I} |0_i\rangle_{II}
	\end{eqnarray}
where $ r  $ is the acceleration setting parameter, $ 0\leq r \leq \pi/4 $ and $ i $ represents the qubit modes.
\\
 Now we discuss the acceleration process in the three cases  mentioned above.
\begin{enumerate}
\item {\it accelerated one qubit\\}
We assume that only the first  qubit $ a $ is accelerated. In this case, thee GHZ state in equation(\ref{4.5}) takes the following form:
\begin{eqnarray}\label{4.8}
	\hat{\rho}^{({acc}_a)}_{GHZ}&=&\mathcal{A}_{11}|000\rangle\langle 000|+ \mathcal{A}_{22}|001\rangle\langle 001|\nonumber\\ &+&\mathcal{A}_{33}|010\rangle\langle 010|+ \mathcal{A}_{44}|011\rangle\langle 011|
\nonumber\\ &+& \mathcal{A}_{55} |100\rangle\langle 100|+
\mathcal{A}_{66}|101\rangle\langle 101|\nonumber\\
&+& \mathcal{A}_{77}|110\rangle\langle 110|+ \mathcal{A}_{88}|111\rangle\langle 111|\nonumber\\
&+&\mathcal{A}_{18}|000\rangle\langle 111|+\mathcal{A}_{81}|111\rangle\langle 000|,
\nonumber\\
\end{eqnarray}
where
\begin{eqnarray}
	\mathcal{A}_{11}&=&\frac{1+3\nu}{8}\cos^2 r, 
\nonumber\\
\quad \mathcal{A}_{22}&=&\frac{1+3\nu}{8}\sin^2 r+\frac{1-\nu}{8}, \qquad
\nonumber\\
\mathcal{A}_{33}&=&\mathcal{A}_{55}=\mathcal{A}_{77}=\frac{1-\nu}{8}\cos^2 r,
\nonumber\\
 \mathcal{A}_{44}&=&\mathcal{A}_{66}=\frac{1-\nu}{8}(\sin^2 r+1),
\nonumber\\
 \mathcal{A}_{88}&=&\frac{1+3\nu}{8}+\frac{1-\nu}{8} \sin^2 r, \quad
 \nonumber\\
 \mathcal{A}_{18}&=&\mathcal{A}_{81}=\frac{\nu}{2}\cos r.
\end{eqnarray}
Inserting Eq(\ref{4.8}) into Eq.(\ref{4.4}), we obtain the Wigner function for accelerated qubit $ a $ as:
\begin{eqnarray}\label{4.9}
	W^{({acc}_a)}&=&\frac{1}{16}\Bigr[(\sqrt{3} \bigr(6 \nu \sin ^3 \theta\cos r \cos 3 \phi
\nonumber\\
&&+\cos \theta  \sin ^2 r (3 \nu \cos 2 \theta +3 \nu+2)\bigl)
\nonumber\\
&&+6 \nu (\cos ^2 \theta  \cos 2 r + \cos 2 \theta+1) +2\Bigl]
\nonumber\\
	\end{eqnarray}

\begin{figure}
	\centering
	\includegraphics[width=0.49\linewidth, height=3.5cm]{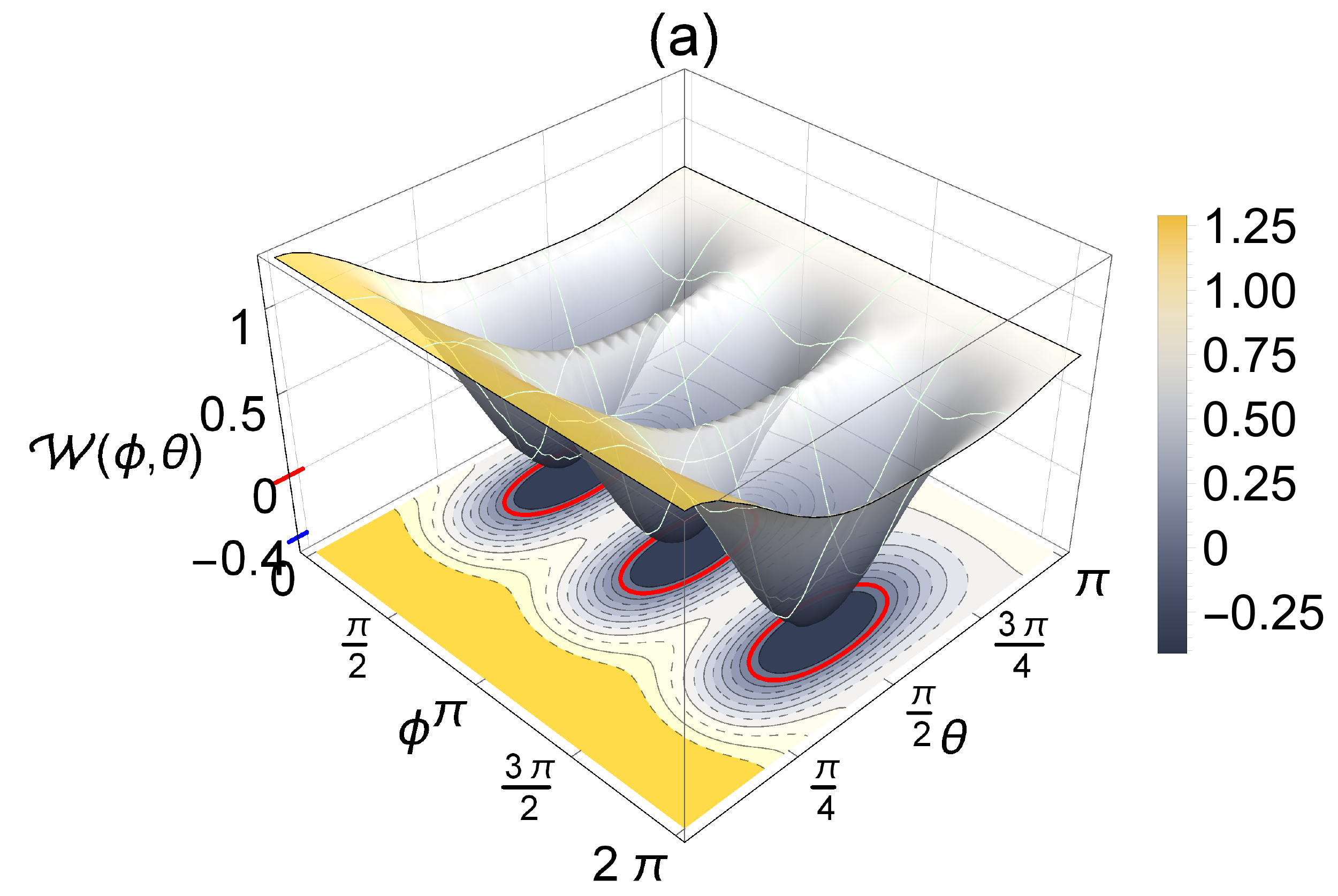}
	\includegraphics[width=0.49\linewidth, height=3.5cm]{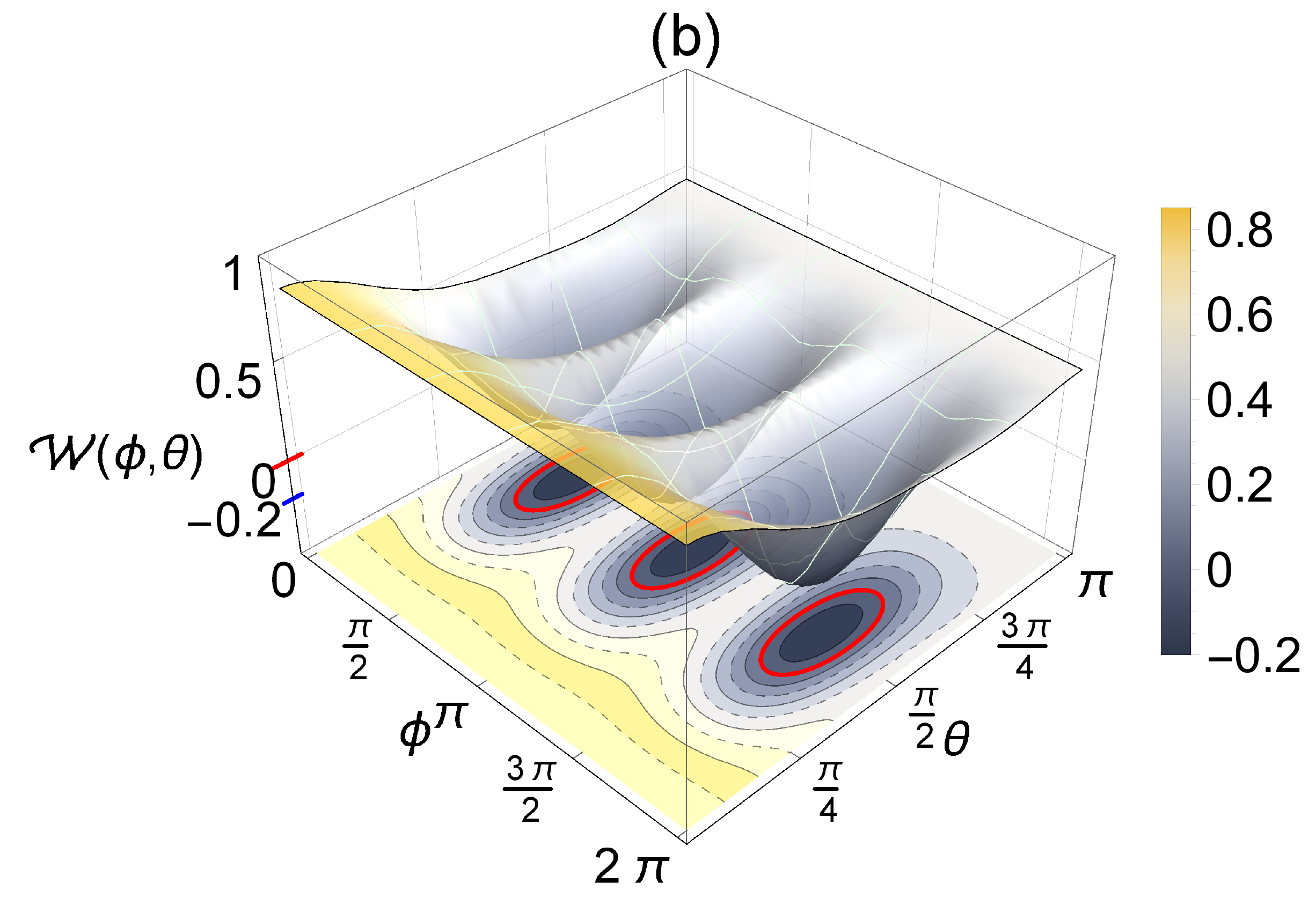}		
    \includegraphics[width=0.9\linewidth, height=4cm]{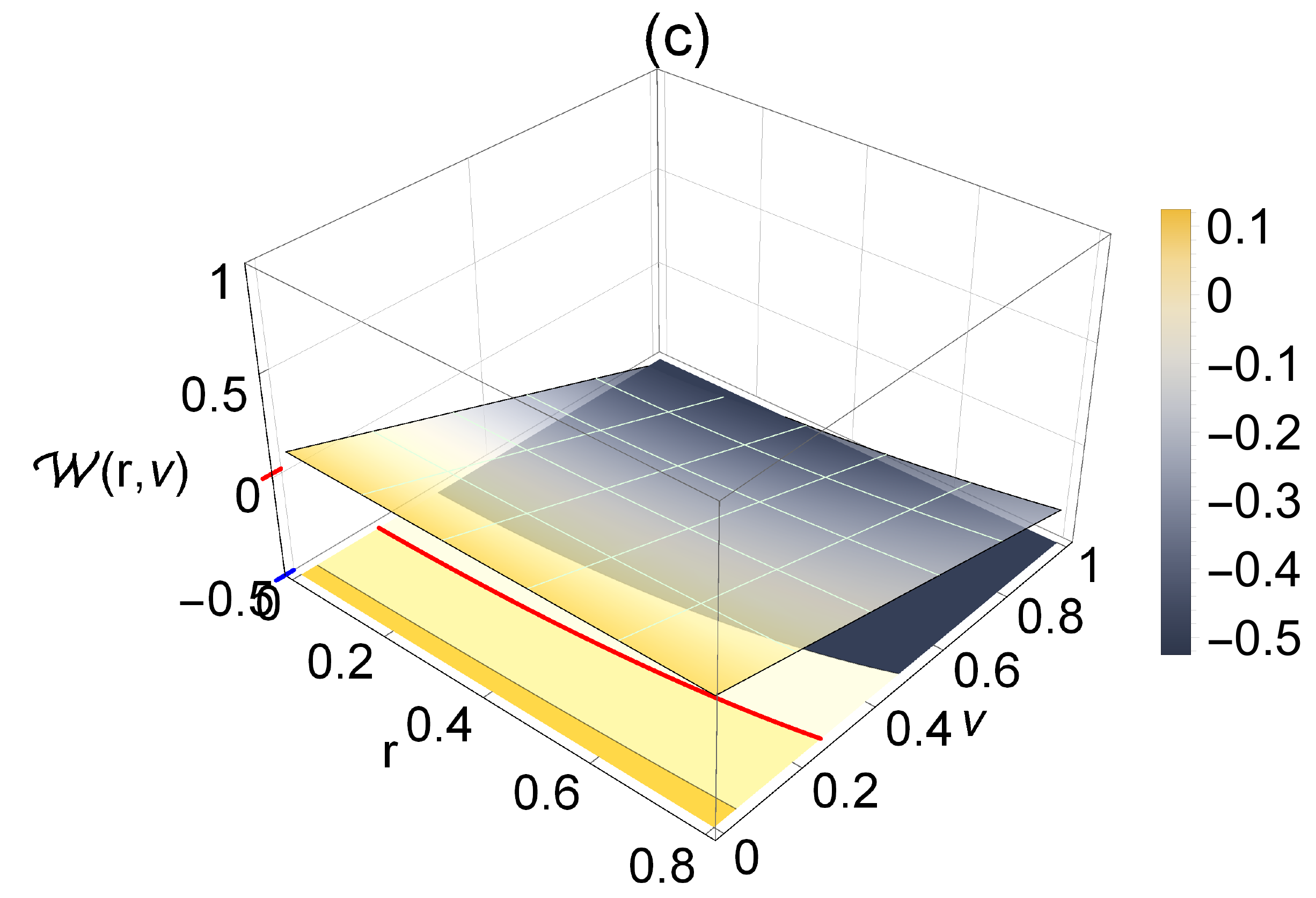}
	\caption{The Wigner function $ W^{acc_{a}}(\phi,\theta)$, $r=0.6,  \phi=\pi $,(a) $ \nu=1$,(b) $\nu=0.3$ and plot $ W(r,\theta) $ and 
 (c)   $ W^{acc_{a}}(\phi=\pi, \theta=\pi/2)$ in the plan $(\nu,~r) $. }
	\label{fig:4.2}
\end{figure}
In Fig.(2), we investigate the effect of the acceleration on the behavior of the Wigner function, where it is assumed that only the first qubit $a$ is accelerated, with $r=0.6$.  It is clear that, the minimum bounds of the Wigner function  are smaller than that displayed in Fig.(1a), for the non-accelerated GHZ state.  Meanwhile, the upper bounds are larger than those displayed in Figs.(1a,1b). However, the quantum correlation is predicted at the same  intervals of $\theta$ and $\phi$ as that displaced for  the non-accelerated case.
Fig.(2c) shows the behavior of $W(\theta,\phi)$ in the plane $(r-\nu)$ at fixed values of $\theta=
\pi/2$, $\phi=\pi$. It is clear that $W(\theta,\phi)$ decreases gradually as $\nu$ increases, namely the quantum correlation increases. However  $W(\theta,\phi)>0$ for  any value $\nu\leq 2$ and arbitrary acceleration.

\item{\it Accelerated two qubit\\}
Now let us consider the the first two qubits $a$ and $b$ are, meanwhile the third qubit "c"  stays statuinary in the inertial frame. By tracing the modes in the second region $II$, the   final accelerated state in the first region $ I $  may be written as,
\begin{eqnarray}\label{4.10}
\hat{\rho}^{({acc}_{ab})}_{GHZ}&&=\mathcal{B}_{11}|000\rangle\langle 000|+ \mathcal{B}_{22}|001\rangle\langle 001|\nonumber\\
&&+\mathcal{B}_{33}|010\rangle\langle 010|+ \mathcal{B}_{44}|011\rangle\langle 011|
\nonumber\\
&&+\mathcal{B}_{55}|100\rangle\langle 100|+\mathcal{B}_{66} |101\rangle\langle 101|\nonumber\\
&&+\mathcal{B}_{77}|110\rangle\langle 110|+\mathcal{B}_{88}|111\rangle\langle 111|
 \nonumber\\
 &&+ \mathcal{B}_{18}|000\rangle\langle 111|+\mathcal{B}_{81}|111\rangle\langle 000|,
 \nonumber\\
\end{eqnarray}
where
\begin{equation*}
\begin{split}
&\mathcal{B}_{11}=\mathcal{A}_{11}\cos^2 r, \quad
\mathcal{B}_{22}=\mathcal{B}_{33}=\mathcal{A}_{22}\cos^2 r,
\\&
 \mathcal{B}_{55}=\mathcal{A}_{33}\cos^2 r, \qquad \mathcal{B}_{44}=\tan ^4 r \mathcal{B}_{11}+\mathcal{A}_{44},
\\&
\mathcal{B}_{66}= \mathcal{A}_{44} \cos^2 r,\quad \mathcal{B}_{88}=\mathcal{A}_{44}(\sin^2 r+1)
\\&
 \mathcal{B}_{77}=\mathcal{A}_{44}\cos^2 r+\frac{1-\nu}{8} \sin^2 r,
\\&
\mathcal{B}_{18}=\mathcal{A}_{18}\cos r.
\end{split}
\end{equation*}

\begin{figure}
	\centering
	\includegraphics[width=0.49\linewidth, height=3.5cm]{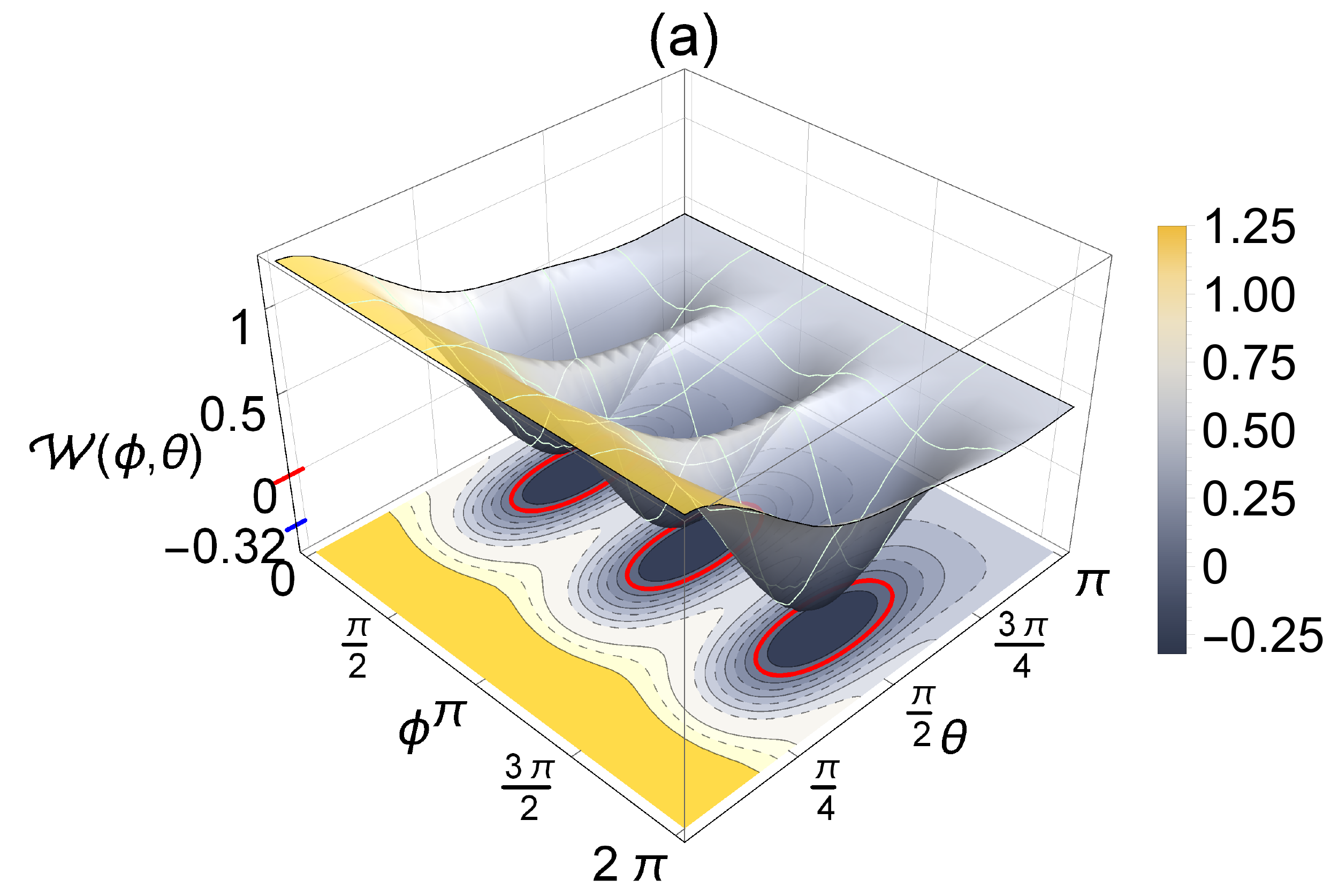}
	\includegraphics[width=0.49\linewidth, height=3.5cm]{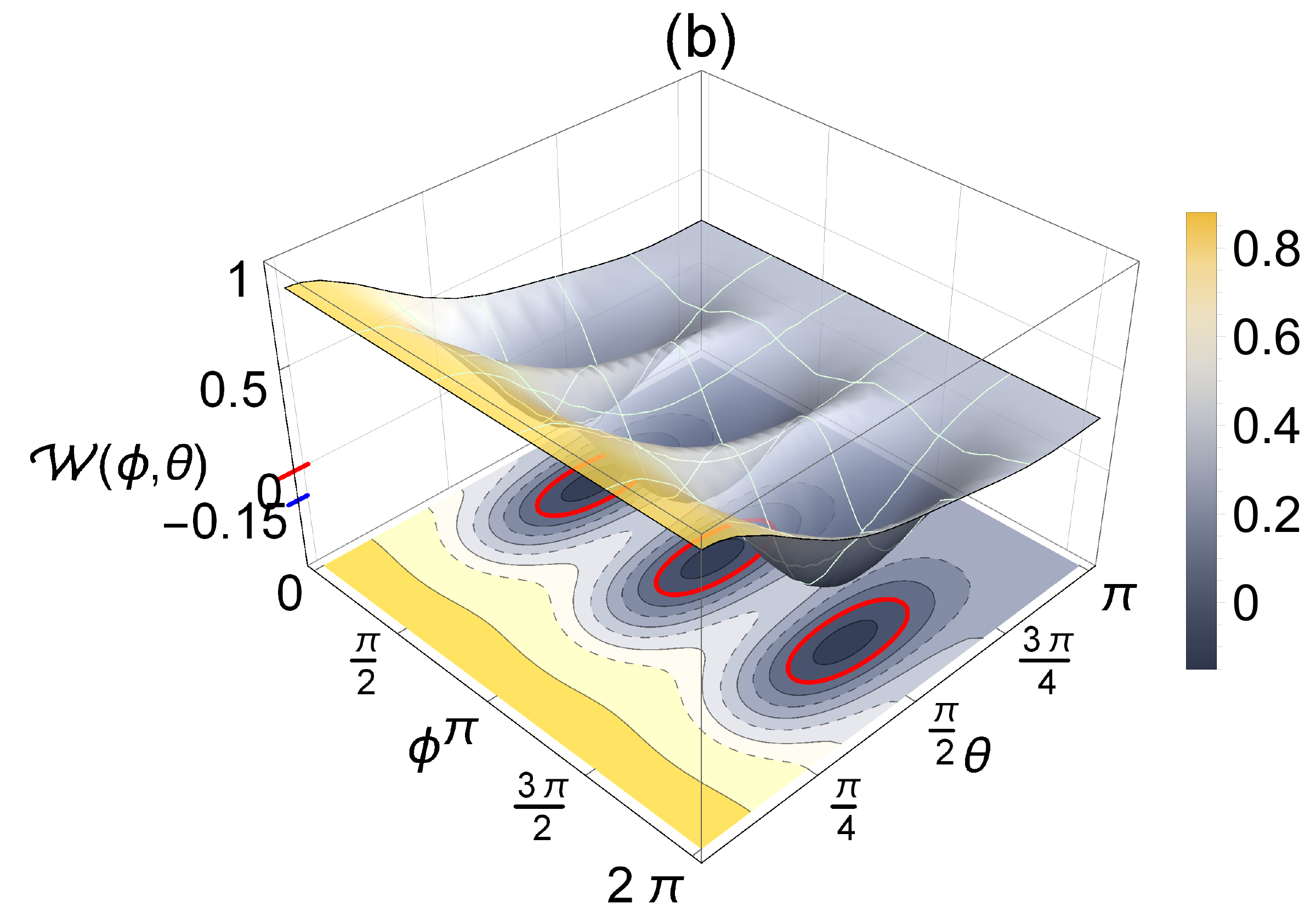}		
    \includegraphics[width=0.8\linewidth, height=4cm]{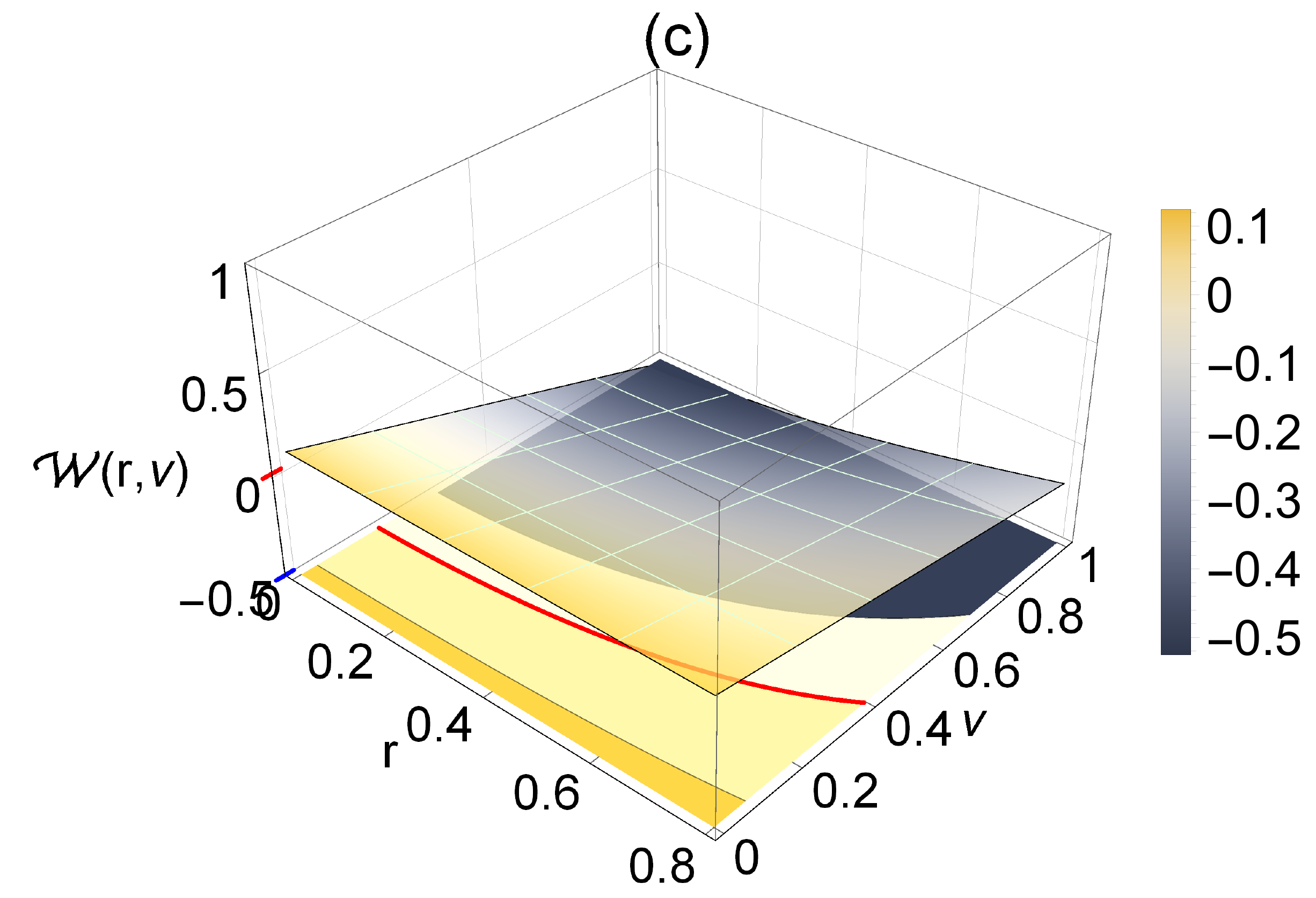}
	\caption{The same as Fig.(2), but it is assumed that two qubits are accelerated}
	\label{fig:4.3}
\end{figure}
By compensation from Eq.(\ref{4.10}) into Eq.(\ref{4.4}), we obtained the Wigner function for both accelerated subsystems "a" and "b" as form:
\begin{eqnarray}\label{4.11}
W^{(ab)}&=&\frac{1}{128}\Bigl[25+48 \sqrt{3} \nu \sin ^3 \theta \cos ^2 r \cos 3 \phi
\nonumber\\
&&+ 9 \cos 2 \theta +33 \nu( \cos 2 \theta +1
\nonumber\\
&&+4 \sqrt{3} \cos \theta  \bigl(3 \nu \cos 2 \theta  \sin ^2 2 r
\nonumber\\
&&+\sin ^2 r (6 \nu \cos 2 r+6 \nu+8)\bigr)
\nonumber\\
&& + \cos ^2\theta  (4 (3 \nu-1) \cos 2 r
\nonumber\\
&&+(\nu+1) \cos 4 r)\Bigr]
\end{eqnarray}
Figs.(3) displays the behavior of the Wigner function when only two qubits are accelerated with the same acceleration. It displays similar features that are predicted in Figs(2).  However, the minimum values that are depicted in Fig.(3a), (3b) are larger than those displayed in Fig.(2a),(2b). This means that the lost quantum correlations that are displayed  by the negativity of the Wigner function, $W(\theta,\phi)$  depends on the numbers of accelerated qubits. The behavior of Wigner in the plan $(r-\nu)$, shows that decreases as $\nu$ increases. As it is shown in Fig.(3c), the upper bounds of negativity are displayed at large values of $\nu$ compared with those are  displayed in Fig.(2c). The positivity of $W(\theta,\phi)$  depicted the existence of the classical correlation on the accelerated GHZ state.  The classical correlation are displayed at any value of $r$ and smaller values of $\nu\in[0,~0.4]$

\item {\it accelerated the three qubits\\}

Finally we  assume that thet hree qubits are accelerated , then in the Rindler space  the final accelerated GHz state is given by

\begin{eqnarray}\label{4.12}
\hat{\rho}^{({acc}_{abc})}_{GHZ}&&=\mathcal{C}_{11}|000\rangle\langle 000|+ \mathcal{C}_{22}|001\rangle\langle 001|\nonumber\\
&&+\mathcal{C}_{33}|010\rangle\langle 010|+
 \mathcal{C}_{44}|011\rangle\langle 011|
 \nonumber\\
 &&+\mathcal{C}_{55}|100\rangle\langle 100|+\mathcal{C}_{66}|101\rangle\langle 101|\nonumber\\
&&+\mathcal{C}_{77}|110\rangle\langle 110|+\mathcal{C}_{12}|111\rangle\langle 111|
\nonumber\\
&&+\mathcal{C}_{18}|000\rangle\langle 111|+\mathcal{C}_{81}|111\rangle\langle 000|,
\end{eqnarray}
where
\begin{equation*}
	\begin{split}
		&\mathcal{C}_{11}=\mathcal{A}_{11}\cos^4 r,~
		\mathcal{C}_{88}=3 \mathcal{A}_{44}\sin^2 r+\frac{1+3\nu}{8}(\sin^6 r+1),
		\\&
		\mathcal{C}_{22}=\mathcal{C}_{33}=\mathcal{C}_{55}=\mathcal{A}_{22}\cos^4 r,\\& \mathcal{C}_{44}=\mathcal{C}_{66}=\mathcal{C}_{77}=\mathcal{A}_{22}\cos^2 r+2\mathcal{A}_{33}\sin^2 r\\&
		\mathcal{C}_{18}=\mathcal{C}_{81}=\mathcal{A}_{18}\cos^2 r.
	\end{split}
\end{equation*}
For this case, the  Wigner function is given By:

\begin{eqnarray}\label{4.13}
W^{(abc)}&=& \frac{48}{128} \Bigr\{\sqrt{3} \nu \sin ^3 \theta \cos ^3 r \cos 3 \phi
\nonumber\\
&-&\left(\frac{3}{2} \eta_+  \cos ^4 r\right)\kappa_1
-\left(6  \eta_- \eta_+^2 \cos ^2 r \right)\kappa_2
\nonumber\\
& +&2\eta_+^3 \left(\sin ^2 r+1\right)\kappa_3
-2  \mu_+ \eta_-^3 \cos ^6 r\Big\}
\nonumber\\
\end{eqnarray}
where $\eta_{\pm}= \left(\sqrt{3} \cos \theta \pm 1\right) $, $ \mu_{\pm}=1\pm3\nu$,\quad
$\kappa_1=(3 \cos 2 \theta +1) (\mu_+ \cos 2 r-\nu-3)$
 $\kappa_2= (\mu_+ \sin ^4 r+2 (1-\nu) (\sin ^2 r+1)$, and \quad $\kappa_3=\mu_+ \sin ^4 r+2\mu_- \sin ^2 r+\mu_+$

\begin{figure}
	\centering
	\includegraphics[width=0.49\linewidth, height=3.55cm]{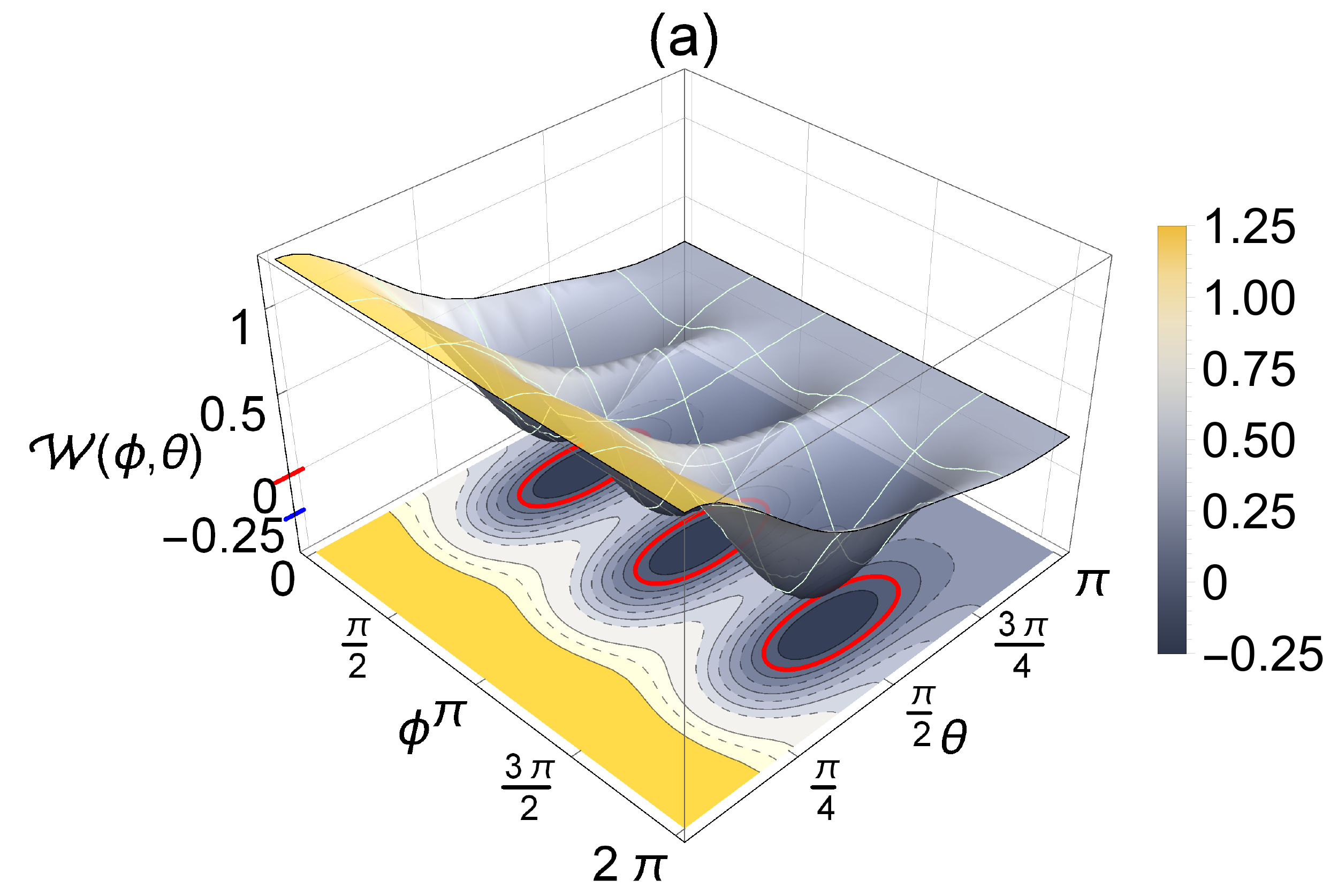}
	\includegraphics[width=0.49\linewidth, height=3.5cm]{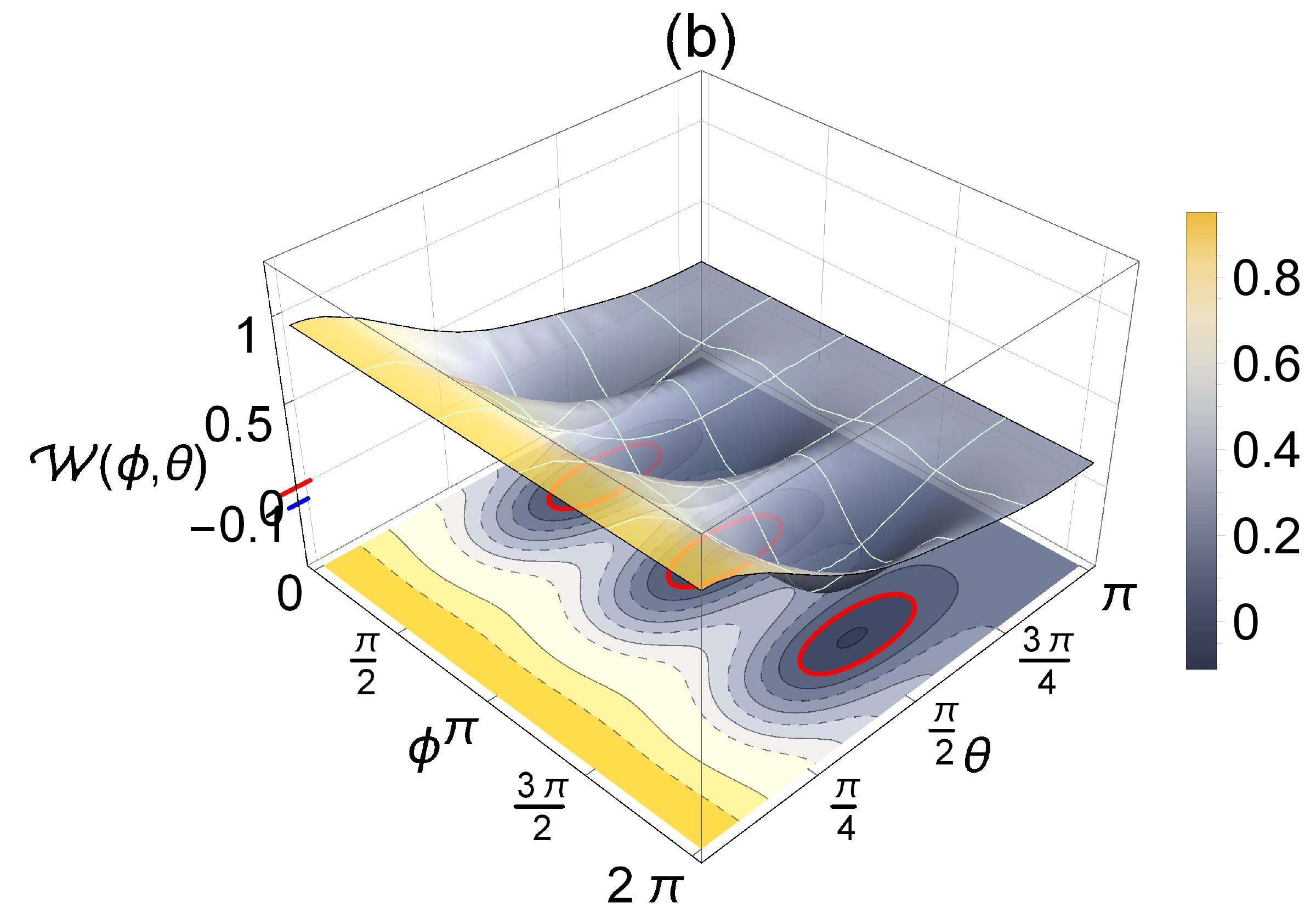}		
    \includegraphics[width=0.8\linewidth, height=4cm]{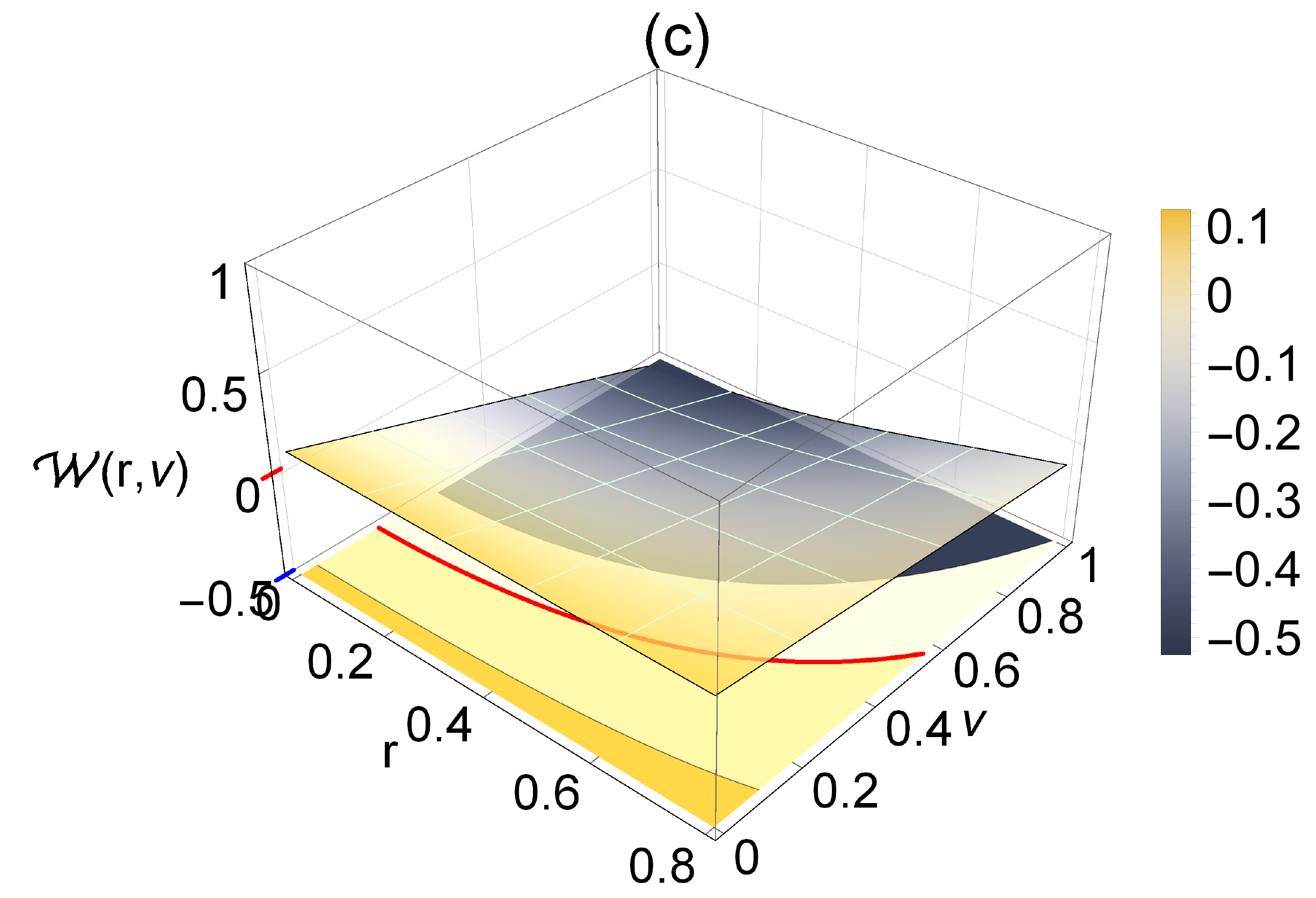}
	\caption{The same as Fig.(1), but when all the three  qubits are accelerated}
	\label{fig:4.4}
\end{figure}
The  behavior of the classical and quantum correlations that are  depicted by Wigner function, $W(\theta,\phi)$  when the three qubits are accelerated is  displayed in Fig.(4). The behavior is similar to that displayed in Fig.(2) and (3), but the classical correlations  increase in the expanses of the quantum correlations.  Moreover, $W(\theta,\phi)$ decreases as the acceleration increases and the mixing parameter $\nu$ decreases.

\begin{figure}
	\centering
	\includegraphics[width=0.49\linewidth, height=4cm]{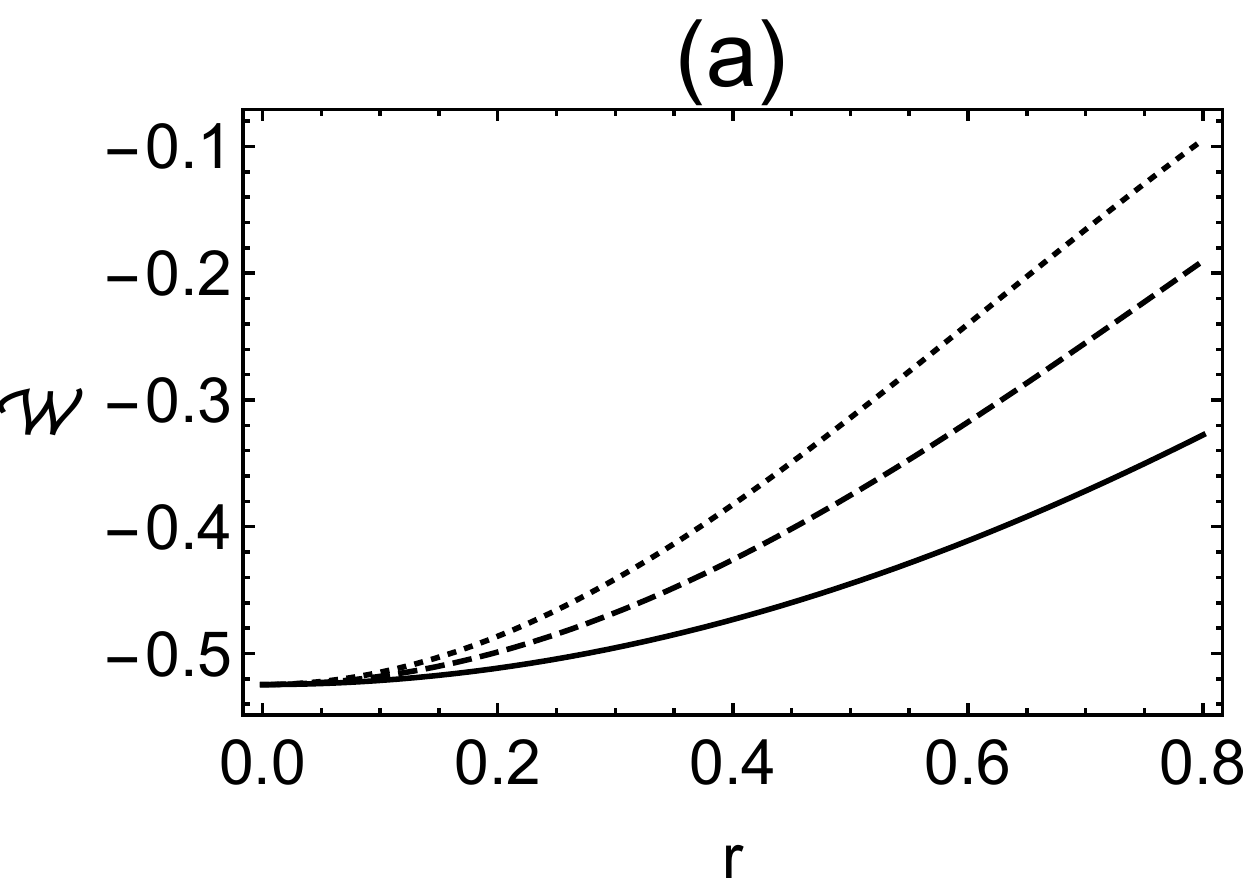}
	\includegraphics[width=0.49\linewidth, height=4cm]{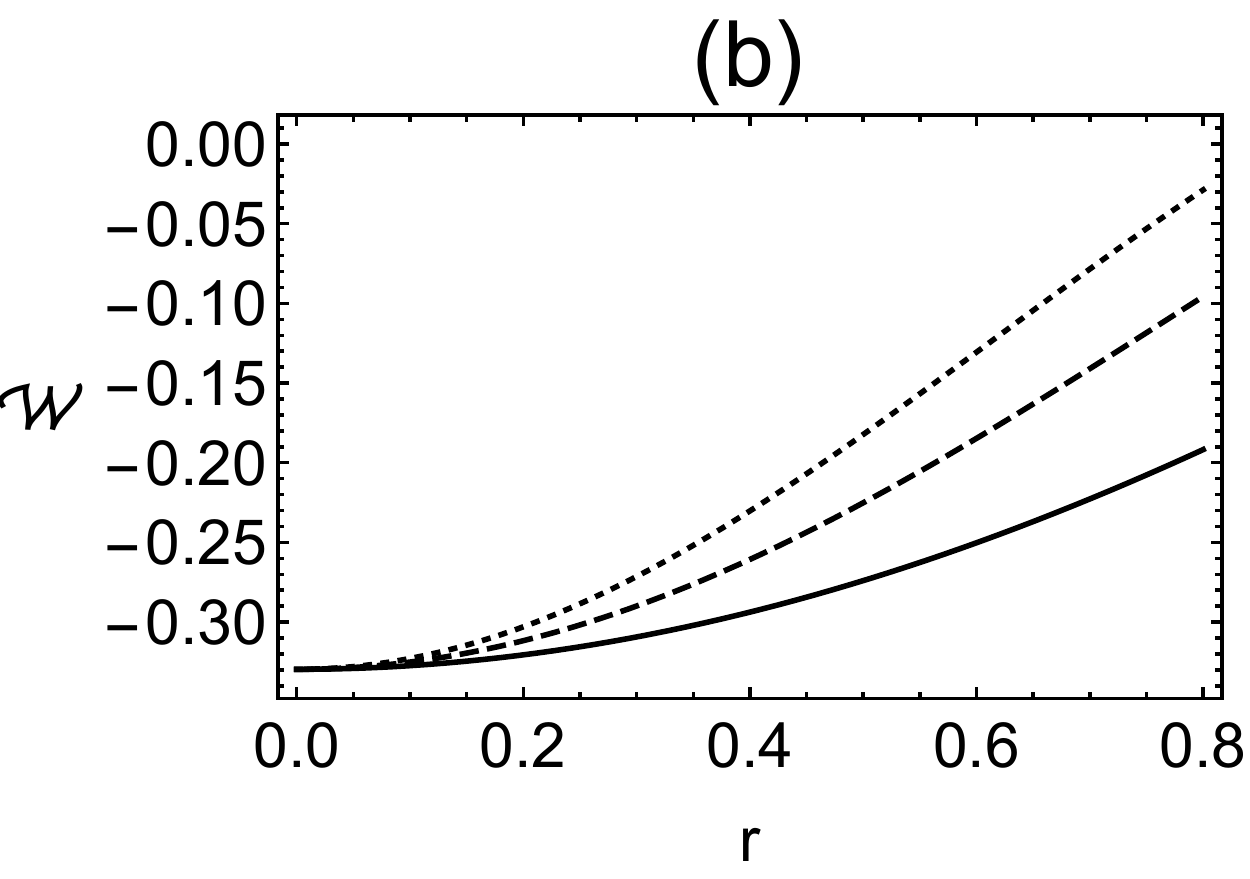}
	\includegraphics[width=0.49\linewidth, height=4cm]{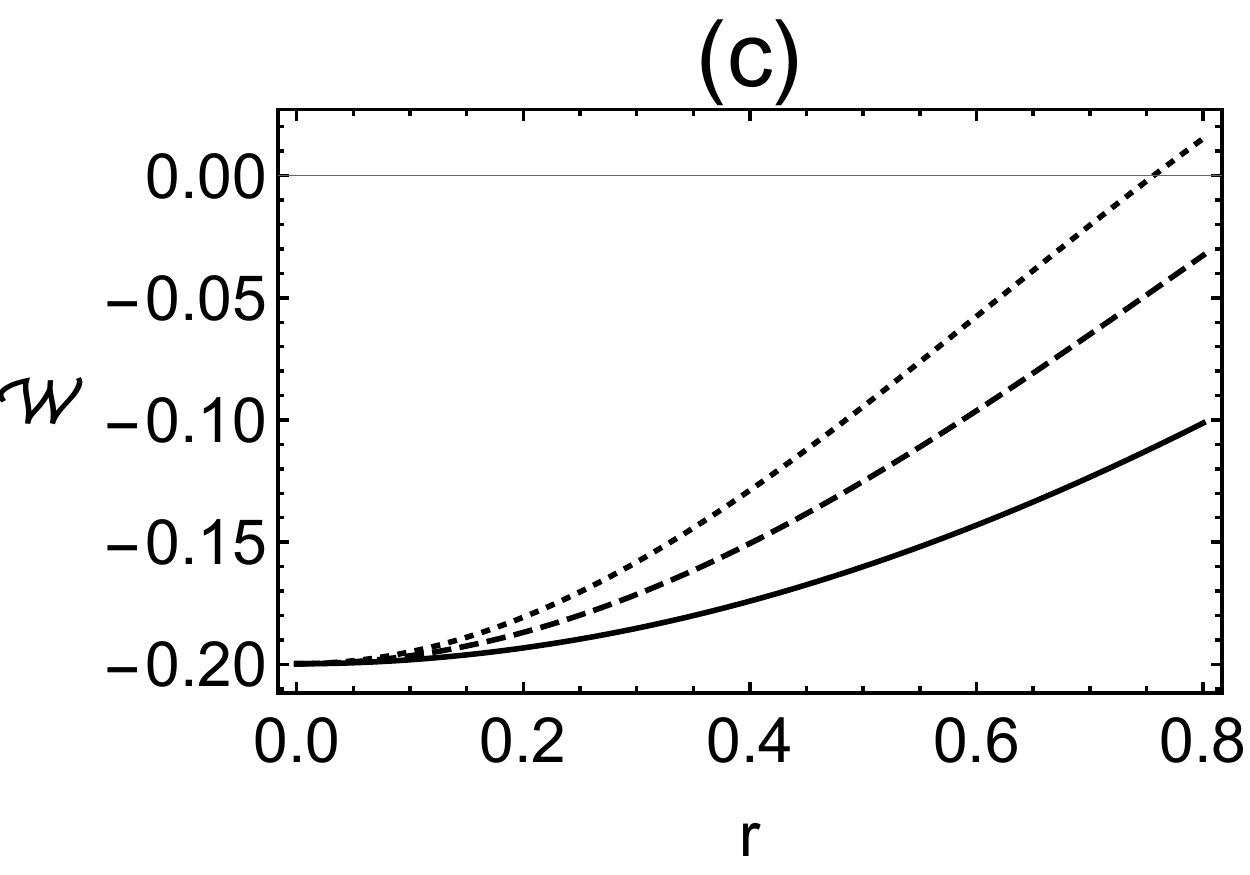}
	\includegraphics[width=0.49\linewidth, height=4cm]{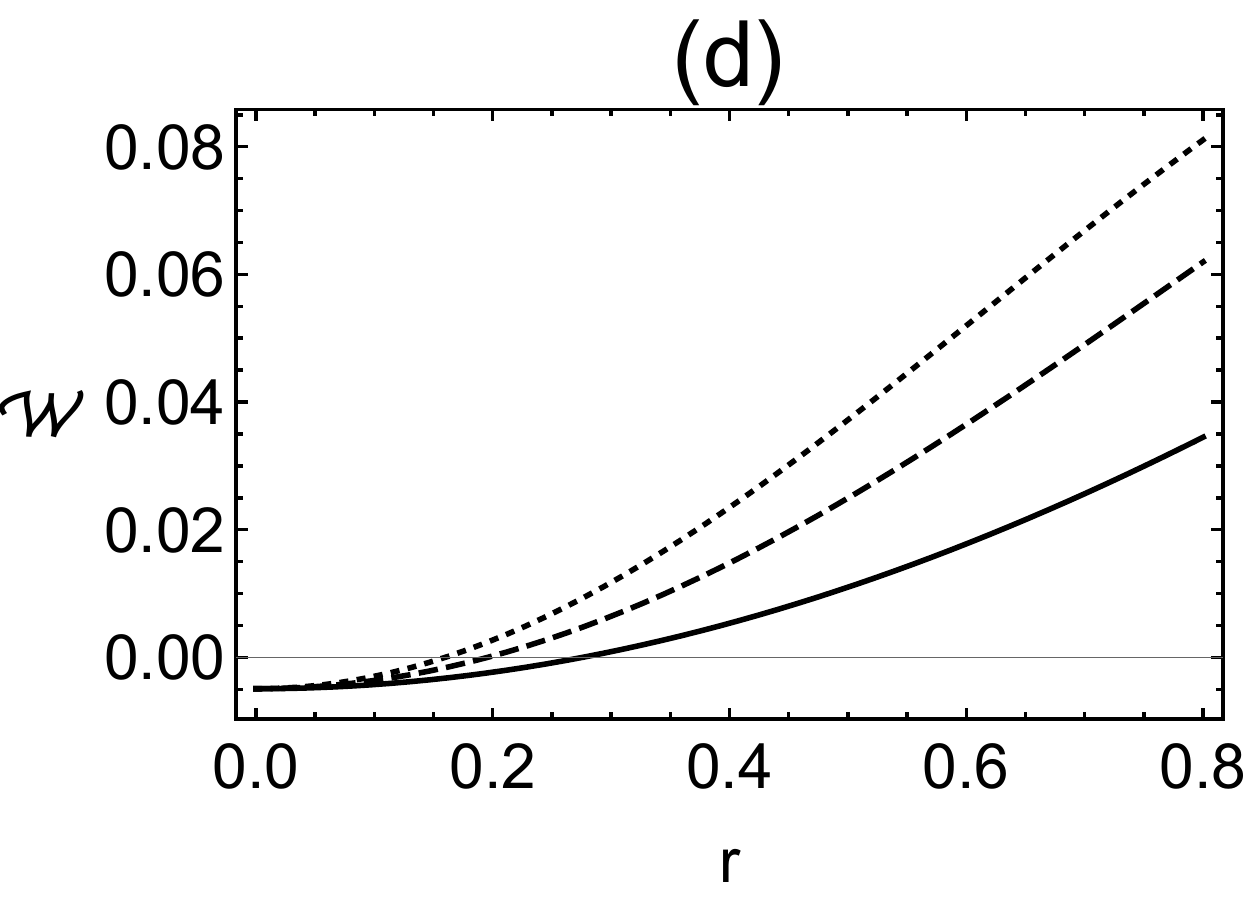}
	\caption{Wigner function against the acceleration parameter for, accelerated one qubit (solid curve), two qubit (dash curve),and three qubit (dot curve), where (a) $ \nu=1$ ,(b) $\nu=0.7 $, (c)$ \nu=0.5 $, and  (d)$ \nu=0.2  $.}
	\label{fig:4.5}
\end{figure}
 In Fig.(5), we  investigate the behavior of $W(\theta=\pi/2,\phi=\pi,r)$ at different values of the mixing parameter $\nu$, where at these distribution angles, Wigner function depicts only the quantum correlation, namely $W<0$.  It is clear that, for  the pure GHZ state, namely $\nu=1$, the quantum quantum correlations are robust against the decoherence due to the acceleration.  However, at  $\nu\neq 0$, namely GHZ state is no longer pure state,  the classical correlations appear and increase on the expense of the quantum correlation.  These  phenomena are clearly seen  from Figs.(5a-5d), where the negativity of W increases as $\nu$ decreases i.e, the purity of GHZ decreases.  Although,  the lose of the  quantum correlation increases as the number of accelerated qubit increases,  the accelerated GHZ still entangled state.

\end{enumerate}

 \section{conclusion}
 The Wigner function of an  accelerated and non-accelerated  GHZ state is discussed. It is shown that, for the non-accelerated GHZ state, the minimum values of Wigner function are centered  around  $\theta=\pi/2$ and different values of $\phi$. The  minimum bounds are shown at maximum values of the mixing  parameter, namely the initial state is a pure GHZ.  However, in the plane of $(\nu,~\theta)$, the negativity of the Wigner function is displaced  at  $\nu>0.2$ and $\theta\in[\pi/4,3\pi/4]$.  The positivity (negativity) of the Wigner function indicates the existence of the classical/quantum  correlations. This confirms  tha the t Wigner function, may be used as an indicators of the entangled and separable behavior of GHZ state.

  For the accelerated GHZ state,  the maximum/minimum bounds of the   Wigner function depend on the distribution angles $(\theta,~\phi)$, the mixing parameter $(\nu)$ and the value of the acceleration $(r)$. It is shown that, in the plan $(\theta,\phi)$ and fixed values of the mixing and acceleration parameters,  the minimum peaks of the Wigner function behane simillarly as WIgner function of the non-accelerated case, namely  they are  centered regularly around $\theta=\pi/2$ and different values of $\phi$. The minimum bounds of the Wigner function are displayed at larger values of the mixing parameter, which indicates the existence of large quantum correlations. Due to the acceleration and the mixing parameter, the classical correlations are depicted by the positive behavior of the Wigner function.  In the plan $(r-~\nu)$ and fixed values of $\theta=\pi/2$ and $\phi=\pi$, i.e.,  only quantum correlations are predicted, the Wigner function decays gradually as the mixing parameter increases. The decay rate depends on the numbers of the accelerated qubits and the value of the mixing parameter.

Moreover, the behavior of the classical correlations are not predicted  the accelerated cases. However,  their appearance due to the non-purity of the initial state. In addition to the classical correlations that contained on the pure state, but don't predicted by the behavior of the Wigner function.

\bibliographystyle{unsrt}
\bibliography{bm}
\end{document}